\documentclass[journal]{IEEEtran}

\usepackage{amsmath}
\usepackage[caption=false,font=footnotesize]{subfig}
\usepackage{cite}
\usepackage{graphicx}
\usepackage{subfig}
\usepackage{balance}
\usepackage{array}
\usepackage{booktabs}
\usepackage{color, soul}
\usepackage{flushend}
\begin{document}
%
\title{Fixed-State Log-MAP Detection for Intensity-\\Modulation and Direct-Detection Optical Systems over Dispersion-Uncompensated Links}
%
%

\author{Shuangyue~Liu,~
        Ji~Zhou,~
        Haide~Wang,~
        Mengqi~Guo,~
        Yueming Lu,~
        and~Yaojun~Qiao
\thanks{This work was supported in part by National Key Research and Development Program of China under Grant 2018YFB1802300, in part by National Natural Science Foundation of China under Grant 61871044 and Grant 61771062, in part by the BUPT Excellent Ph.D. Students Foundation under Grant XTCX201813, in part by the Open Fund of IPOC (BUPT) (No. IPOC2019A001),  and in part by Natural Science Foundation of Guangdong Province under Grant 2019A1515011059. \emph{(Corresponding author: Ji Zhou and Yaojun Qiao.)}}
\thanks{S. Liu, M. Guo, and Y. Qiao are with the State Key Laboratory of Information Photonics and Optical Communications, School of Information and Communication Engineering, Beijing University of Posts and Telecommunications (BUPT), Beijing 100876, China (e-mail: liusy\_moon@bupt.edu.cn; guomengqi@bupt.edu.cn; qiao@bupt.edu.cn).}
\thanks{J. Zhou and H. Wang are with the Department of Electronic Engineering, College of Information Science and Technology, Jinan University, Guangzhou 510632, China (e-mail: zhouji@jnu.edu.cn; 1834041007@stu2018.jnu.edu.cn).}
\thanks{Y. Lu is with the Key Laboratory of Trustworthy Distributed Computing and Service, Ministry of Education, Beijing University of Posts and Telecommunications (BUPT), Beijing 100876, China (e-mail: ymlu@bupt.edu.cn).}
\thanks{}}

\markboth{}%
{}
%

\maketitle

\begin{abstract}
In this paper, an optimized detection based on log-maximum a posteriori estimation with the fixed number of surviving states (fixed-state Log-MAP) is proposed to cooperate with equalizers to deal with the spectral distortions caused by limited bandwidth and chromatic dispersion for intensity-modulation and direct-detection (IM/DD) optical systems. The equalizers compensates the spectral distortions and optimized detection decodes the useful bits from the noise. For accurately extracting the bits from more serious noise, the optimized detection with larger memory length is required. However, the classical optimized detection such as maximum likelihood-sequence estimation (MLSE) requires exponential-growing computational complexity and storage with the increasing memory length. The fixed-state Log-MAP detection can decrease the computational complexity and storage from the exponential order to linear order. Therefore, the fixed-state Log-MAP detection can compensate more distortions compared to MLSE under the same hardware condition. We experimentally verify the fixed-state Log-MAP detection in a C-band 64 Gbit/s IM/DD on-off keying optical system over a 100 km dispersion-uncompensated link. Under the same hardware condition, the fixed-state Log-MAP detection has a 2 dB improvement of receiver sensitivity compared to MLSE. In conclusion, the fixed-state Log-MAP detection shows the potential for practical IM/DD optical systems.
\end{abstract}

\begin{IEEEkeywords}
Fixed-state Log-MAP detection, intensity-modulation and direct-detection, chromatic dispersion, spectral distortions.
\end{IEEEkeywords}

%
\IEEEpeerreviewmaketitle
\section{Introduction}
\label{sec:intr}
%
%
%
%
\IEEEPARstart{W}{ith} the rapid development of bandwidth-thirsty services such as 4K/8K high-definition video, interactive games and social media applications, the traffic of data center interconnects is increasing dramatically \cite{Zhong:review,DCI:review,Zhou:DCI}. Owing to low-cost, low-power-consumption and small-footprint characteristics, optical interconnects tend to adopt the intensity-modulation and direct-detection (IM/DD) 
optical systems\cite{2019OFC:IMDD,IMDD,coherent}, which is recommend by P802.3cn Task Force for 400G Ethernet within $40$ km \cite{400GZR}. However, due to the square law detection for IM/DD optical systems, chromatic dispersion (CD) of optical fiber causes severe power fading on signal spectrum \cite{powerfading,ECOC:CD,Tang:2020OFC}. With an increase in the capacity-distance product, the distortions become increasingly more severe, which rapidly degrades the performance of system \cite{Tang:CD}. For the higher symbol rate for the next-generation 800G Ethernet, the transmission distance is so limited that it is difficult to reach more than 10 km \cite{204OOK2,800G,600G}. Therefore, how to compensate CD-caused distortions is an urgent issue for short-reach optical interconnects.

O-band transmission essentially circumvents CD-caused distortions, but has a high fiber loss and lacks mature optics \cite{2020OFC:OBand}. Therefore, many studies focus on compensating the CD-induced power fading for C-band IM/DD transmission. Optical dispersion compensation using dispersion compensation fiber (DCF) is the simplest way, but DCF introduces a high fiber loss and high implementation cost \cite{DCF}. Electrical dispersion compensation including CD pre-compensation \cite{CDPre}, single sideband or vestigial sideband modulation (SSB/VSB) \cite{SSB,VSB}, and Kramers-Kronig receiver \cite{KK}, can effectively compensate CD-caused distortions. However, a complex transceiver structure and extra expensive devices will increase the implementation cost and footprint. \cite{compare}. Therefore, advanced digital signal processing (DSP) techniques without changing the traditional structure of IM/DD optical systems are researched to alleviate CD-caused impairments. Popular DSP includes Tomlinson-Harashima precoding (THP) \cite{THP1,THP2,THP3}, decision feedback equalizer (DFE) \cite{ARDFE,Tang:2019OFC,Tang:PAM480km}, and maximum likelihood sequence estimation (MLSE) \cite{Zhou:56G100km,MLSE}. Our previous work proposed adaptive channel-matched detection (ACMD) for dispersion-uncompensated links \cite{Zhou:64G100km}. To realize the optimal detection, MLSE in ACMD requires a large memory length to accurately decode the useful bits from serious noise. However, the bottleneck of MLSE lies in an exponential-growing computational complexity and storage with an increase in memory length, placing a heavy burden on the computing resources and internal storage space of the hardware. Generally, due to the limit of compute capability, the realizable memory length of MLSE is $\le 15$ for on-off keying (OOK) systems \cite{Zhou:56G100km}. 

\begin{figure*}[!t]
\centering
\centering{\includegraphics[width=0.9\linewidth, height=6.2cm]{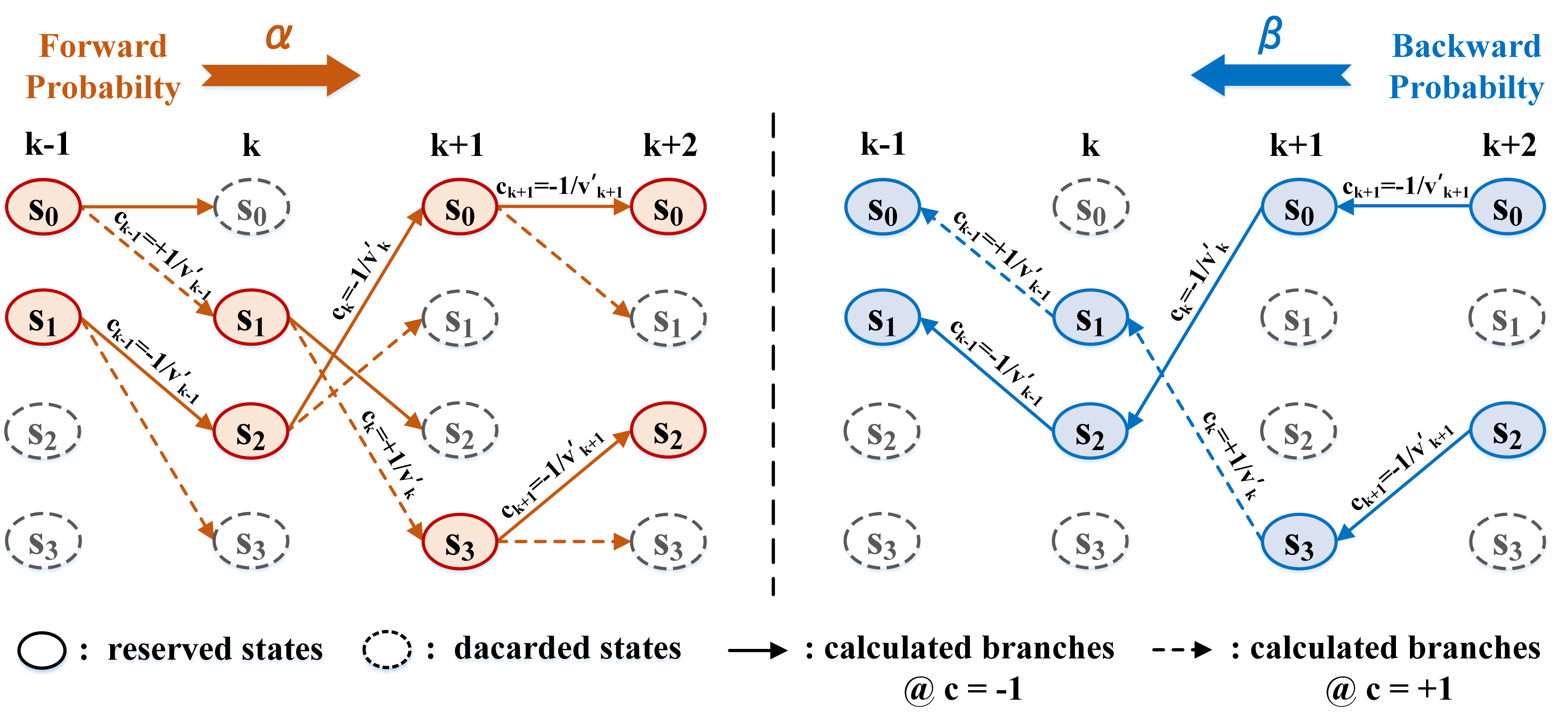}}
\caption{Trellis diagram of fixed-state Log-MAP detection algorithm. (States: $s_0=(-1,-1)$, $s_1=(-1,+1)$, $s_2=(+1,-1)$, $s_3=(+1,+1)$, and $M=2$ states are reserved. $c_k/v'_k$ on the branch denotes the input/output).}
\label{Fig1}
\end{figure*}

In this paper, an optimized detection based on log-maximum a posteriori estimation with the fixed number of surviving states $M$ (fixed-state Log-MAP) is proposed to replace MLSE to solve the restricted memory length. The fixed-state Log-MAP detection is experimentally demonstrated in a C-band 64 Gbit/s IM/DD OOK system over a 100 km dispersion-uncompensated link. The main contributions of this paper are summarized as follows:
\begin{itemize}
  \item  We propose a fixed-state Log-MAP detection with a low computational complexity and a less storage space. When the memory length is $L$, the complexity order decreases from $\mathcal{O}(2^{L+1})$ of MLSE to $\mathcal{O}(2M)$ of fixed-state Log-MAP detection, and the state storage decreases from $2^{L}$ states to $M$ states.
  \item Experimental results show that the fixed-state Log-MAP detection availably overcomes the constraint of memory length. Under the same hardware condition, the receiver sensitivity using fixed-state Log-MAP detection has a 2 dB improvement than that using MLSE at 7\% hard-decision forward error correction (HD-FEC) limit.
\end{itemize}

The rest of this paper is organized as follows. The principles are analyzed in Section \uppercase\expandafter{\romannumeral2}. In Section \uppercase\expandafter{\romannumeral3}, the experimental setup and DSP are presented. In Section \uppercase\expandafter{\romannumeral4}, we analyze and discuss the experimental results. Finally, this paper is concluded in Section \uppercase\expandafter{\romannumeral5}.

\section{Principle}
At the receiver side, the equalizers using polynomial nonlinear equalizer (PNLE) $\&$ DFE are applied to compensate the channel distortions, including limited-bandwidth-induced high-frequency distortions and CD-induced spectral nulls. However, the equalizers inevitably enhance the in-band noise with different enlargement factors for different frequencies. The enhanced in-band noise leads to the limited signal to noise ratio (SNR) of the equalizers output, resulting in the undesirable transmission performance. The noise is colored in the output of the equalizers. The optimal detection is proved in the system with additive white Gaussian noise channel \cite{DSPBook}. Therefore, for accurately extracting the useful bits from serious noise, the noise-whiten post-filter followed by fixed-state Log-MAP detection is adopted.

\subsection{Noise-whiten post-filter}
The noise-whiten post-filter shapes the signal with a known channel response to suppress the enhanced in-band noise. The output $v_k$ of $(L+1)$-tap noise-whiten post-filter is given by
\begin{equation}
v_k = h_0 \cdot x_{k} + \sum_{i=1}^{i=L} h_i \cdot x_{k-i},
\label{eq0}
\end{equation}
where $x$ is the output of electrical equalizers with the colored noise, and $h_i$ is the tap coefficients. The tap coefficients can be obtained adaptively based on Yule-Walker equations for the autoregressive (AR) coefficients extraction \cite{AR}. The number of taps is $(L+1)$. For transmission links with severe distortions, the post-filter requires many taps to match the channel response. At the right side of Eq. (\ref{eq0}), the first term represents the current symbol at the time $k$, and the second term represents the known inter-symbol interference (ISI) with the memory length of $L$ caused by noise-whiten post-filter. After noise-whiten post-filter, the fixed-state Log-MAP detection is required to decode useful bits from the severe ISI.

\subsection{Fixed-state Log-MAP detection}
The output of fixed-state Log-MAP detection is the log-likelihood ratio (LLR). For a two-level pulse-amplitude modulation (PAM2) signal, the output can be expressed as \cite{log-MAP}
\begin{equation}
LLR \left(u_{k} \right)={\mathrm{ln}}\;{\frac{P \left(u_k=+1|z \right)}{P \left(u_k=-1|z \right)}},
\label{eq1}
\end{equation}
where $u_k$ is the transmitted PAM2 symbol with the elements of $\left\{-1, 1\right\}$. $P(u_k|z)$ is the probability of transmitted symbol $u$ at time $k$ in the condition of the received sequence $z$. $P(u_k|z)$ is derived as 
\begin{equation}
  P(u_k|z) = {\mathrm{exp}} \left[{\mathrm{ln}}(\alpha_{k-1}(s')) + {\mathrm{ln}}(\gamma_k(s',s))+{\mathrm{ln}}(\beta_k(s))\right],
\label{eq2}
\end{equation}
where $s'$ and $s$ represent the state at the previous time and at the next time, respectively. $\gamma_k\left(s',s\right)$, $\alpha_{k-1}\left(s'\right)$, and $\beta_k\left(s\right)$ mean branch transition probability, forward probability, and backward probability, respectively.

The state trellis of the fixed-state Log-MAP detection is depicted in Fig. \ref{Fig1} (supposed that $L=2$). The maximum number of states is $2^{L}=4$. However, some states have a low probability, which are very possible not in the optimal detection path. Even though these low-probability states are discarded, the detection performance may be not affected at all. Therefore, the number of surviving states $M$ is introduced to reduce the computational complexity and state storage. In Fig. \ref{Fig1}, we take $M=2$ as an example. Each reserved state generates $2$ branches, including $c_k=+1$ and $c_k=-1$. We only need to calculate $M$ states $*$ 2 branches / state $=2M$ branches. A branch of the trellis represents that the state $s'$ at time $k$ with the input of $c_{k}$ and the output of $v'_{k}$ is transferred to the state $s$ at time $k+1$. The branch transition probability is given by
\begin{equation}
  {\mathrm{ln}}(\gamma_{k}\left(s',s\right)) = - \frac{1}{2\sigma^2} \cdot \left|z_k - v'_k \right|^2,
\label{eq3}
\end{equation}
in which 
\begin{equation}
v'_k = \sum_{i=0}^{L} h_i c_k, 
\end{equation}
$c_k$ is the PAM2 constellation, and $h_i$ is the channel information provided by noise-whiten post-filter.

During the calculation of forward probability $\alpha$, we reserve the $M=2$ states with the maximum probability and discard the others. The forward probability is calculated recursively as follow:
\begin{equation}
  {\mathrm{ln}}(\alpha_k\left(s\right))={\mathrm{ln}}\sum_{s'} {\mathrm{exp}} \left({\mathrm{ln}}(\alpha_{k-1}\left(s'\right))+{\mathrm{ln}}(\gamma_k\left(s',s\right))\right).
\label{eq4}
\end{equation}

\renewcommand\arraystretch{2}
\begin{table}[!t]
\setlength{\abovecaptionskip}{0.cm}
\setlength{\belowcaptionskip}{-0.cm}
\centering
\caption{Comparison of the computational complexity and state storage between MLSE and fixed-state Log-MAP.}
\begin{tabular}{c|c|c}
 \hline\hline
 Type of algorithm & MLSE & Fixed-state Log-MAP \\ [0.5ex]
 \hline
 State storage & $2^{L}$ states & $M$ states   \\[0.5ex]
 \hline
 Complexity order & $\mathcal{O}(2^{L+1})$ & $\mathcal{O}(2M)$ \\ [0.5ex]
 \hline
 Remarks & Exponential order of $L$ & Linear order of $M$ \\ [0.5ex]
 \hline\hline
\end{tabular}
\label{table:1}
\end{table}

The calculation of backward probability $\beta$ is conducted according to the path reserved by forward probability without any additional path extensions. The backward probability is calculated recursively as follow:
\begin{equation}
  {\mathrm{ln}}(\beta_{k}\left(s'\right))={\mathrm{ln}} \sum_{s} {\mathrm{exp}} \left({\mathrm{ln}}(\beta_{k+1}\left(s\right)) + {\mathrm{ln}}(\gamma_{k}\left(s',s\right))\right).
\label{eq5}
\end{equation}

Finally, the LLR output supports to soft-decision forward error correction (SD-FEC) to obtain more decoding gain. In this paper, we are not focused on forward error correction (FEC). The LLR is sent to a hard-decision module to directly obtain the transmitted symbols. $LLR \left(u_{k} \right)\ge 0$ means that the probability of $u_{k}=+1$ is greater than $u_{k}=-1$. Therefore, the hard-decision output is $u_{k}=+1$ at $LLR \left(u_{k} \right)\ge 0$. Conversely, the hard-decision output is $u_{k}=-1$. The output of hard-decision module is expressed as
\begin{equation}
{u_k} = \left\{ \begin{array}{l}
+1,{\kern 13pt} LLR \left(u_{k} \right)\ge 0\\
-1,{\kern 13pt} LLR \left(u_{k} \right)< 0.
\end{array} \right.
\label{eq6}
\end{equation}

A comparison of the computational complexity and state storage between MLSE and fixed-state Log-MAP detection is shown in Table \ref{table:1}. For MLSE, $2^{L}$ states require to be stored, and the computational complexity is in order of $\mathcal{O}(2*2^{L})$. Both computational complexity and state storage suffer from exponential growth as the memory length increases, resulting in the limitation of memory length. Generally, the maximum memory length of MLSE is $L=15$ owing to the limited compute capability. Otherwise, the running space exceeds running memory resulting in an error. Moreover, the enormous computation and storage place a huge burden on the computing resources and storage resources for hardware implementation. As for fixed-state Log-MAP detection, we reserve $M$ surviving states and calculate the probability of $2M$ branches. The $M$-state storage and $\mathcal{O}(2M)$-order computational complexity are linearly related to surviving states and are independent of memory length. Therefore, the fixed-state Log-MAP detection overcomes the limitations of memory length, making it possible to use larger memory length to more accurately extract the useful information from severe noise.



\section{Experimental setups}
\label{section3}

\begin{figure}[!t]
\centering
\includegraphics[width=0.9\linewidth]{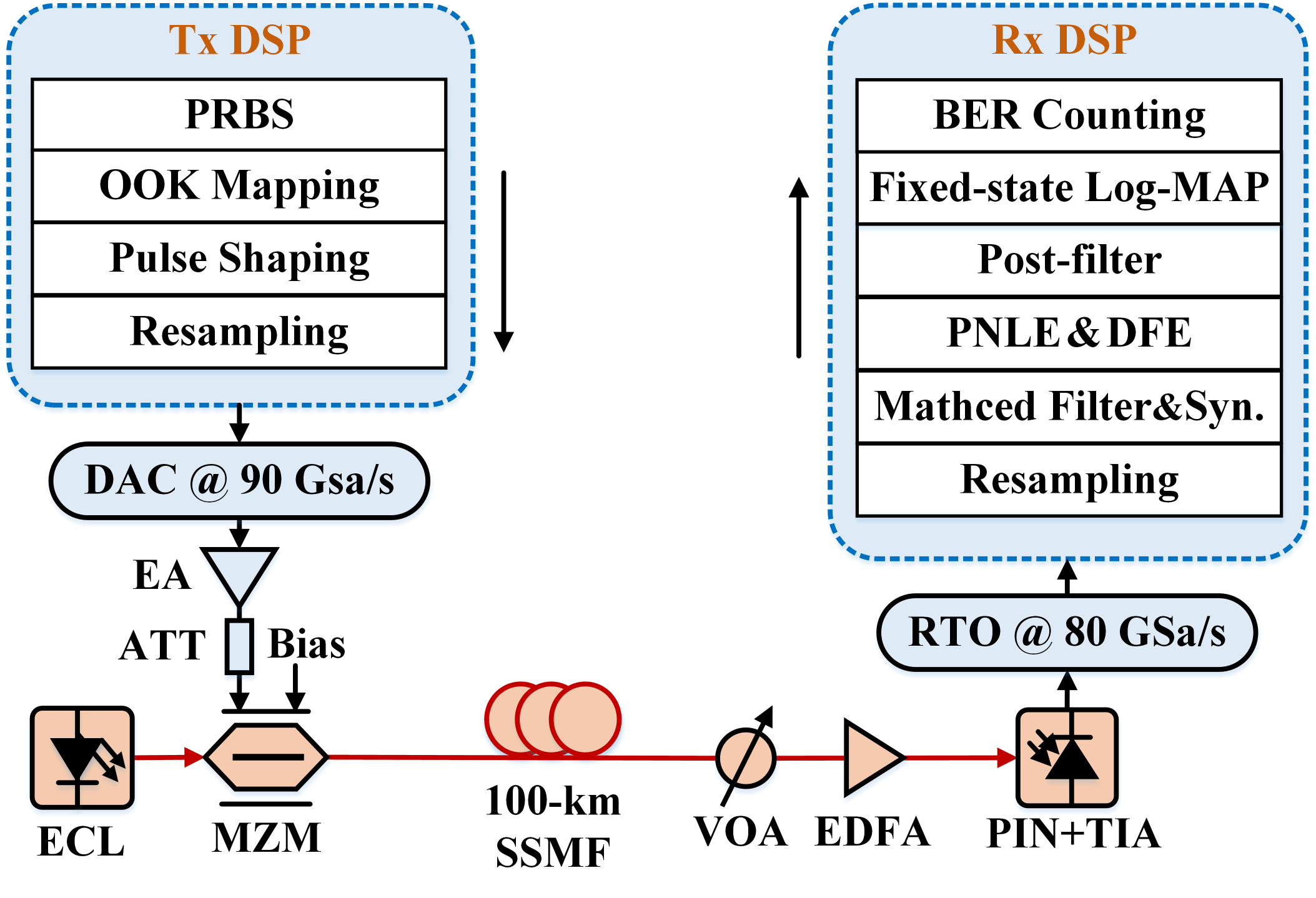}
\caption{Experimental setups of a C-band 64 Gbit/s IM/DD optical OOK system over a 100 km dispersion-uncompensated link.}
\label{Fig2}
\end{figure}

Fig. \ref{Fig2} shows the experimental setups of a C-band 64 Gbit/s IM/DD optical OOK system over a 100 km dispersion-uncompensated link. At the transmitter, the pseudo-random binary sequences (PRBS) are mapped into PAM2 symbols with the elements of $\left\{-1, 1\right\}$. A digital PAM2 frame consists of 5000 training sequences and 77240 payload symbols. The PAM2 signal is added a DC component to generate a unipolar OOK signal. The OOK signal is then up-sampled and shaped by a digital root-raised cosine (RRC) filter. The resampling is then employed for matching the sampling rate of digital-to-analog converter (DAC). The data is uploaded into the 90 GSa/s DAC with an 8 bit resolution and 3 dB bandwidth of 16 GHz. Afterwards, the electrical OOK signal is amplified by a electrical amplifier (EA, Centellax OA4SMM4) followed by a 3 dB attenuator (ATT). The external cavity laser (ECL) with 1550.116 nm center wavelength is used to generate the optical carrier. The electrical OOK signal is modulated on the optical carrier using a 40 Gbps Mach-Zehnder modulator (MZM) at the single-drive mode (Fujitsu FTM7937EZ) with $+2$ V bias voltage, which is to ensure the modulation within the linear regions. The generated optical OOK signal is fed into a 100 km SSMF. The launch power is set to 7 dBm, which is the maximum launch power of the device. The link loss of the 100 km transmission is approximately 20 dB.

 \begin{figure}[!t]
\centering
\centering{\includegraphics[width=2.45in]{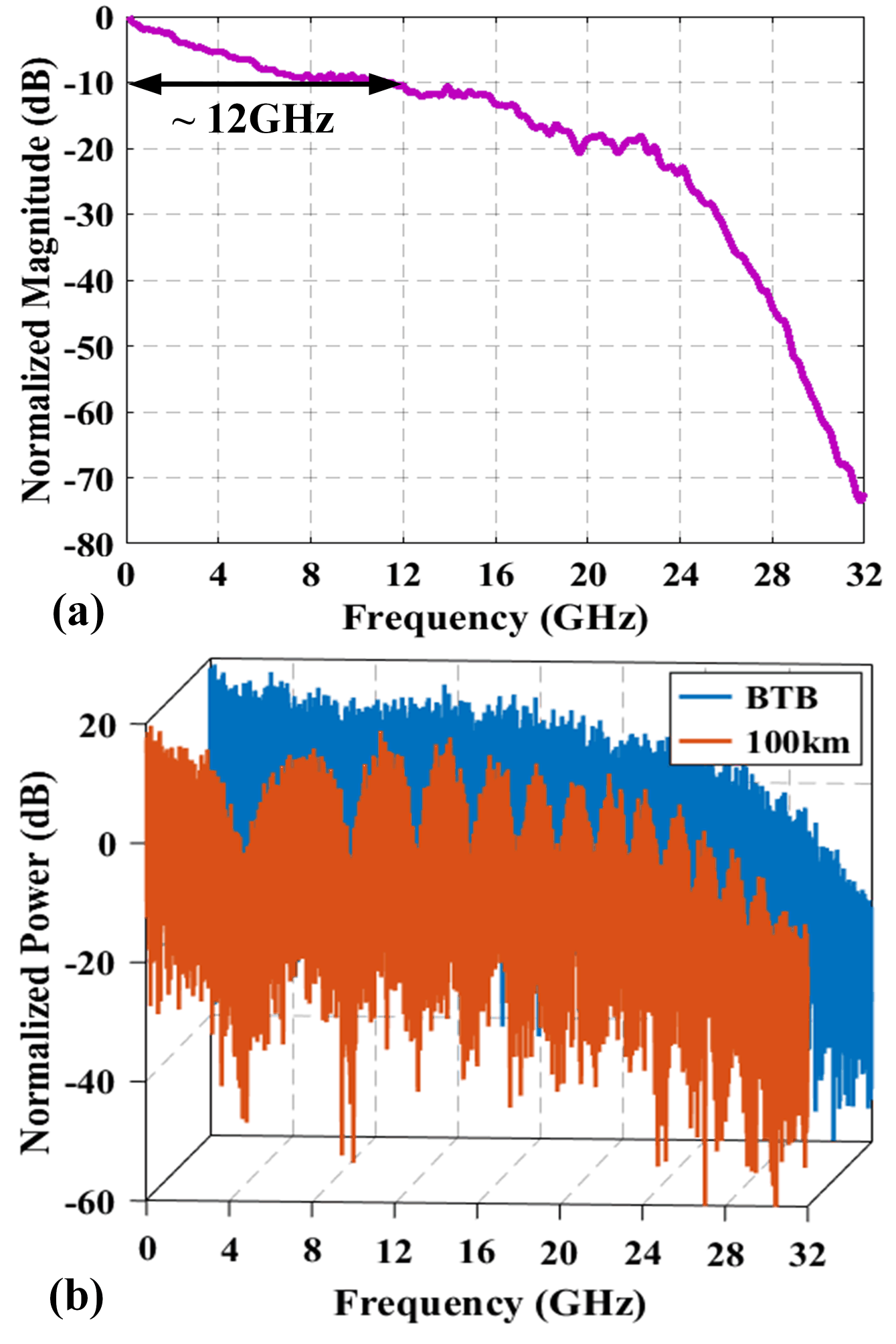}}
\caption{(a). Frequency response of the system for OBTB transmission; (b). Normalized power spectrum of received 64 Gbit/s OOK signal at OBTB and after 100 km SSMF transmission.}
\label{Fig3}
\end{figure}

At the receiver, a variable optical attenuator (VOA) is applied to adjust the received optical power (ROP). The optical signal is then amplified by an Erbium-doped ﬁber ampliﬁer (EDFA), which ensures that the input power of the photoelectric detector is sufficient. The output power of EDFA is fixed at -4 dBm. The optical signal is converted into an electrical signal by a 31 GHz PIN with a trans-impedance amplifier (PIN-TIA) (Finisar MPRV1331A). The electrical signal is then sampled by a 80 GSa/s real-time oscilloscope (RTO) with 36 GHz bandwidth. However, when the bandwidth of RTO ranges from 32 GHz to 36 GHz, the filtering effect is severe. Finally, the off-line DSP is conducted on the sampling electrical signal, including resampling, RRC matched filter, time synchronization, PNLE $\&$ DFE, noise-whiten post-filter, fixed-state Log-MAP detection, and bit error ratio (BER) counting.

\section{Experimental results and analysis}
\label{section4}
\subsection{Frequency response and power spectrum}
Fig. \ref{Fig3} (a) shows the frequency response of the system for optical back to back (OBTB) transmission. The whole system has 3 dB bandwidth of about 2.5 GHz and 10 dB bandwidth of about 12 GHz. The baseband bandwidth of 64 Gbit/s OOK signal is 32 GHz. The channel fading is fast when the frequency is beyond 24 GHz. Fig. \ref{Fig3} (b) shows the normalized power spectrum of received signal at OBTB and after 100 km SSMF transmission. For OBTB transmission, the power spectrum fades at the high-frequency regions due to the limited bandwidth of the system. The signal suffers from severe high-frequency distortions. For a 100 km dispersion-uncompensated link, the power spectrum appears 14 spectral nulls. The reason is that the function of fiber channel in the frequency domain for IM/DD systems is expressed as (supposed that only CD effect of optical fiber is considered for simplicity)
\begin{equation}
  Re{\{H(f)\}}=\cos(2\pi^2\beta_2L_0f^2),
\label{eq7}
\end{equation}
 in which $\beta_2$ denotes the group velocity coefficient, $L_0=100$ km is the fiber length, and $f$ is the signal frequency. The signal is subjected to severe distortions with a cosine function of the quadratic frequency. The CD-induced power fading results in several spectral nulls, which causes severe signal distortions and drops in performance. Moreover, the spectral nulls increase with the increase of transmission distance and signal bandwidth. 
 
\subsection{Performance analysis}
\begin{figure}[!t]
\centering
\includegraphics[width = 8.75cm,height = 4cm]{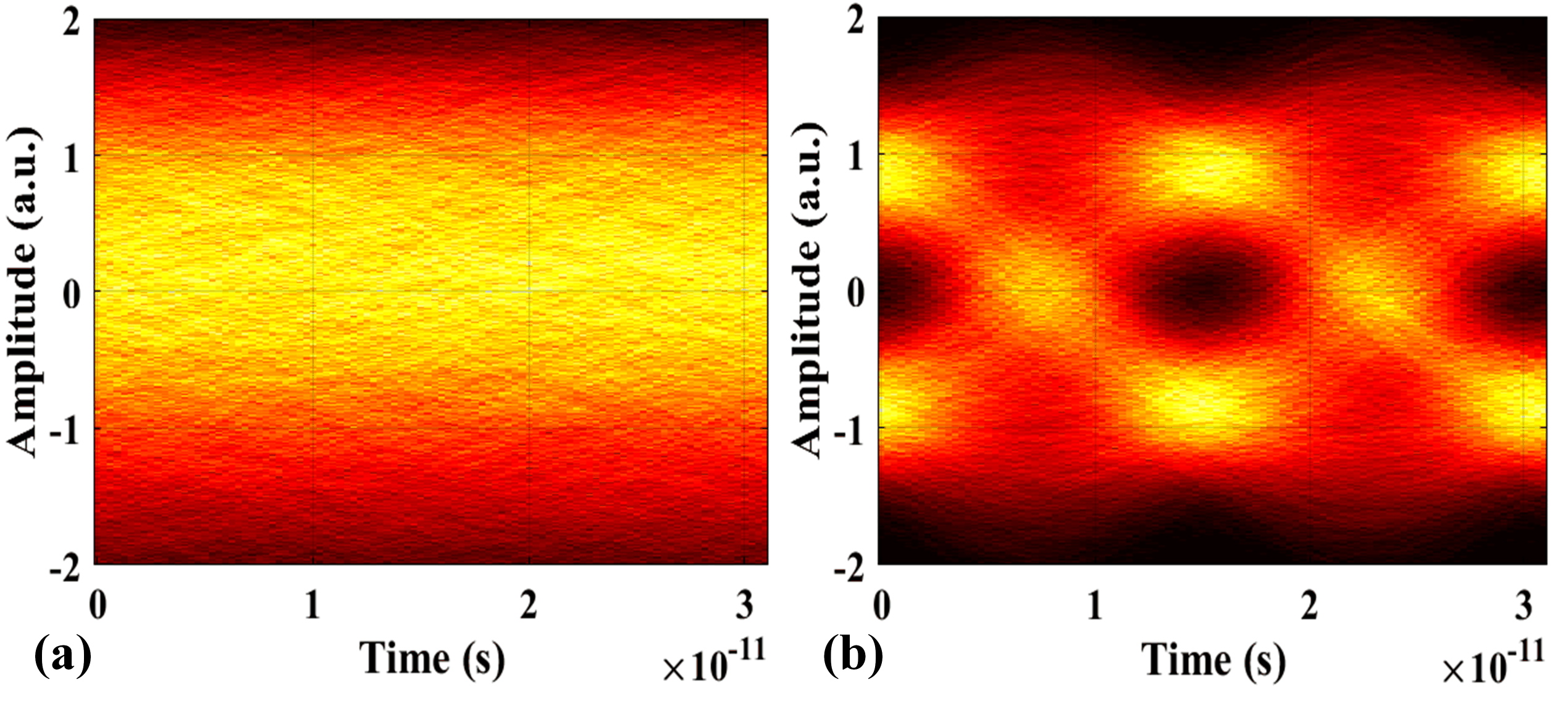}
\caption{Eye diagram of the received 64 Gbit/s OOK signal (100 km SSMF) (a). before PNLE $\&$ DFE; (b). after PNLE $\&$ DFE.}
\label{Fig4}
\end{figure}

\begin{figure}[!t]
\centering
\includegraphics[width=\linewidth]{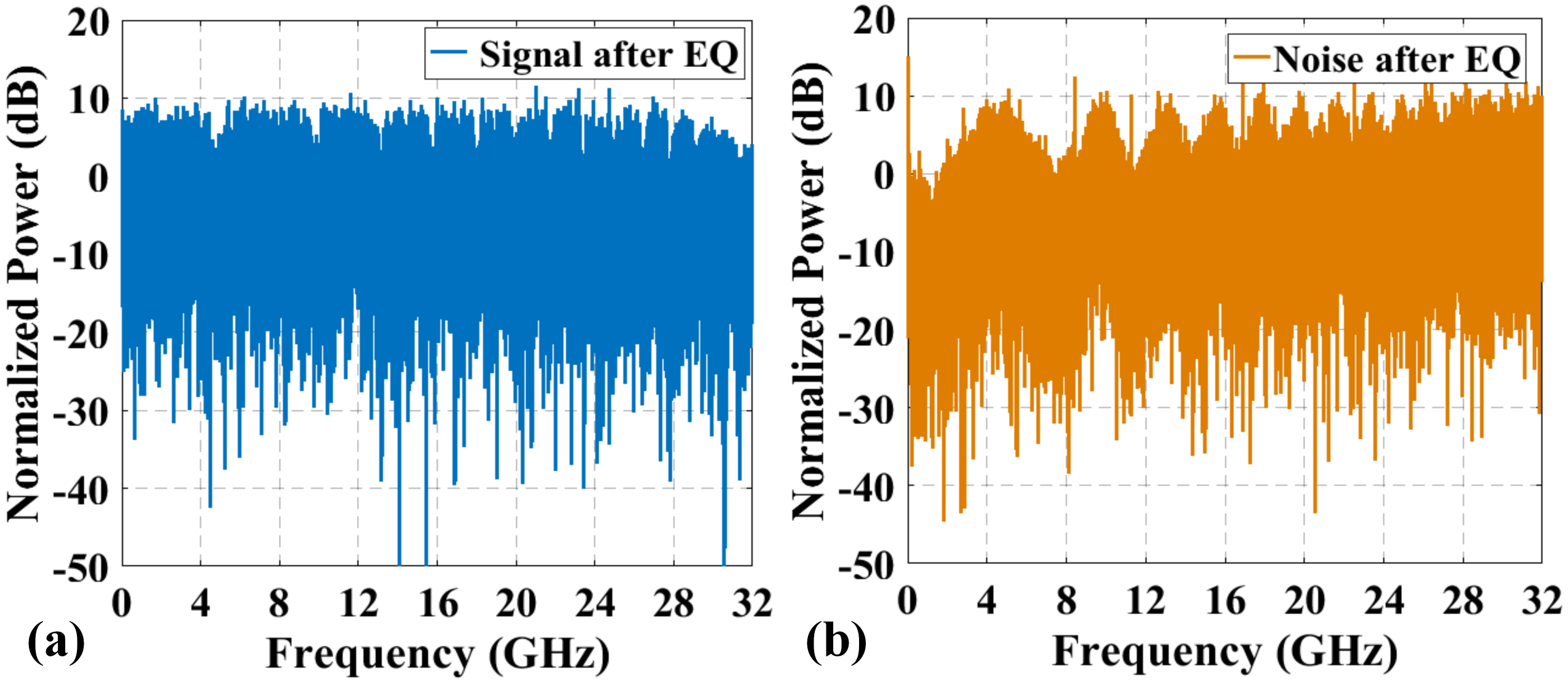}
\caption{Normalized power spectrum of (a). received 64 Gbit/s OOK signal (100 km SSMF) after PNLE $\&$ DFE; (b). noise after PNLE $\&$ DFE.}
\label{Fig5}
\end{figure}

We conducted a 64 Gbit/s optical OOK experiment over a 100 km dispersion-uncompensated link to verify the feasibility of the fixed-state Log-MAP detection. The HD-FEC with $7\%$ overhead is employed, and the net rate is approximately 56.18 Gbit/s ($64$ Gbit/s $\times 77240/82240/ (1+7\%) \approx 56.18$ Gbit/s). Through a trade-off between system performance and computational complexity, the tap number of 3-order PNLE and DFE are set to $(291, 81, 41)$ and $(71, 61)$, respectively. Fig. \ref{Fig4} (a) and (b) illustrate the eye diagram of the received OOK signal before and after PNLE $\&$ DFE, respectively. Before PNLE $\&$ DFE, the eye diagram is very indistinct owing to the severe signal distortions. After PNLE $\&$ DFE, the eye diagram opens. This indicates that PNLE $\&$ DFE can compensate most of channel impairments. However, the eyelids are still thick. This reason is that while the channel distortions are compensated, the in-band noise is also enhanced for the corresponding frequency.

Fig. \ref{Fig5} (a) and (b) illustate the normalized power spectrum of received signal and noise after PNLE $\&$ DFE, respectively. Compared with the normalized power spectrum before PNLE $\&$ DFE (Fig. \ref{Fig3} (b)), most of power-fading effects have been compensated as shown in Fig. \ref{Fig5} (a). However, the in-band noise is enhanced at the frequency of spectral nulls and at the high-frequency regions. The noise becomes colored as shown in Fig. \ref{Fig5} (b), which limits the performance of signal detection. Therefore, a $(L+1)$-tap noise-whiten post-filter is used to whiten in-band noise to obtain optimum SNR. The fixed-state Log-MAP detection with $L$ memory length eliminates the ISI caused by post-filter to realize the optimal detection.

\begin{figure}[!t]
\centering
\includegraphics[width=2.7in]{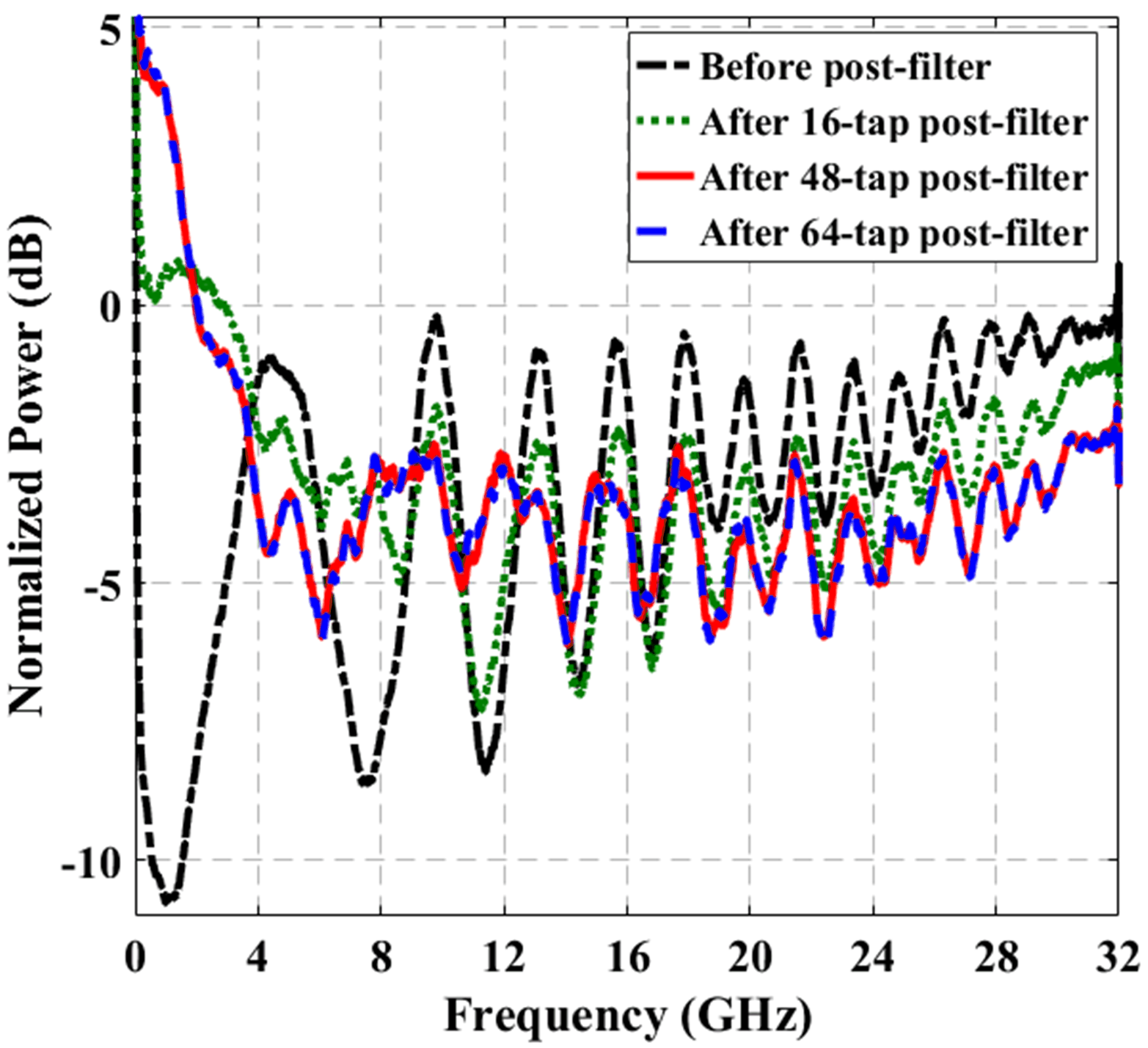}
\caption{Smoothed noise power spectrum before post-filter (black dashed line), after a 16-tap post-filter (green dotted line), after a 48-tap post-filter (red solid line), and after a 64-tap post-filter (blue short line).}
\label{Fig6}
\end{figure}

Fig. \ref{Fig6} depicts the smoothed noise power spectrum before post-filter (black dashed line), after a 16-tap post-filter (green dotted line), after a 48-tap post-filter (red solid line), and after a 64-tap post-filter (blue short line). When a 16-tap post-filter is used, a part of enhanced noise at the frequency of spectral nulls and high-frequency regions is suppressed. After a 48-tap post-filter, the enhanced noise is further suppressed. Moreover, the 48-tap post-filter has the same suppression effect as the 64-tap post-filter.

\begin{figure}[htbp]
\centering
\includegraphics[width=3in]{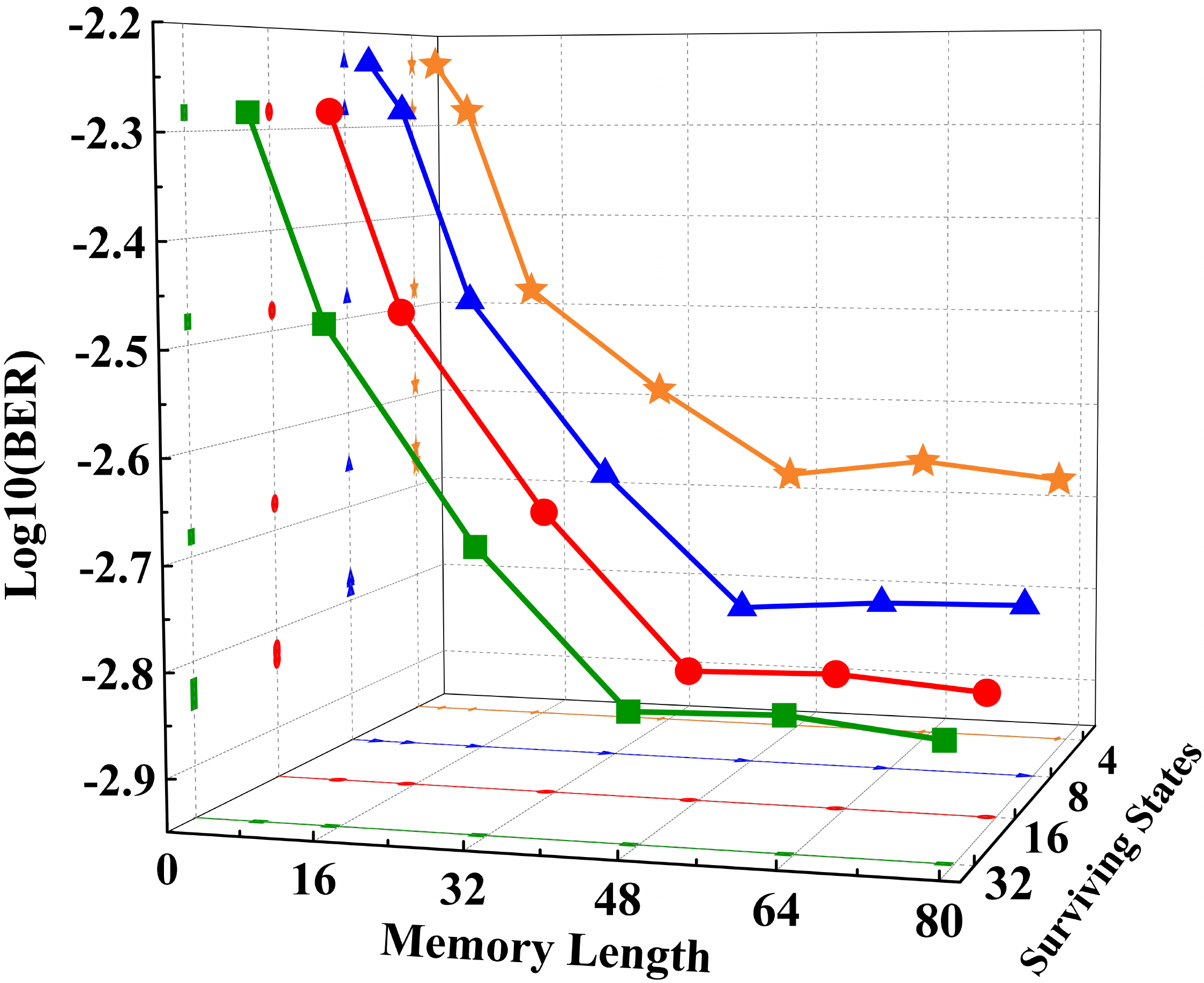}
\caption{BER against memory length and surviving states of fixed-state Log-MAP detection for a 64 Gbit/s IM/DD optical OOK system over a 100 km SSMF transmission.}
\label{Fig7}
\end{figure}
Fig. \ref{Fig7} shows the BER performance against the memory length and surviving states of fixed-state Log-MAP detection. The ROP is set to -14 dBm. The number of surviving states is $M$. Fig. \ref{Fig7} illustrates that the larger $M$ yields a better BER performance since it guarantees a higher probability to reserve the global maximum posterior probability detection. The performance saturates when $M$ is $16$. Besides, with an increase in the memory length, the useful bits is decoded more accurately from serious noise, and BER performance improves gradually. When the memory length of fixed-state Log-MAP detection is set to $47$, the BER performance saturates and realizes the optimal detection. It is worth noting that the performance trend is consistent with the analysis of noise suppression effect in Fig. \ref{Fig6}. Therefore, the optimal memory length is $L=47$, and the number of surviving states of fixed-state Log-MAP is $M=16$.

\begin{figure}[!t]
\centering
\includegraphics[width=3.2in]{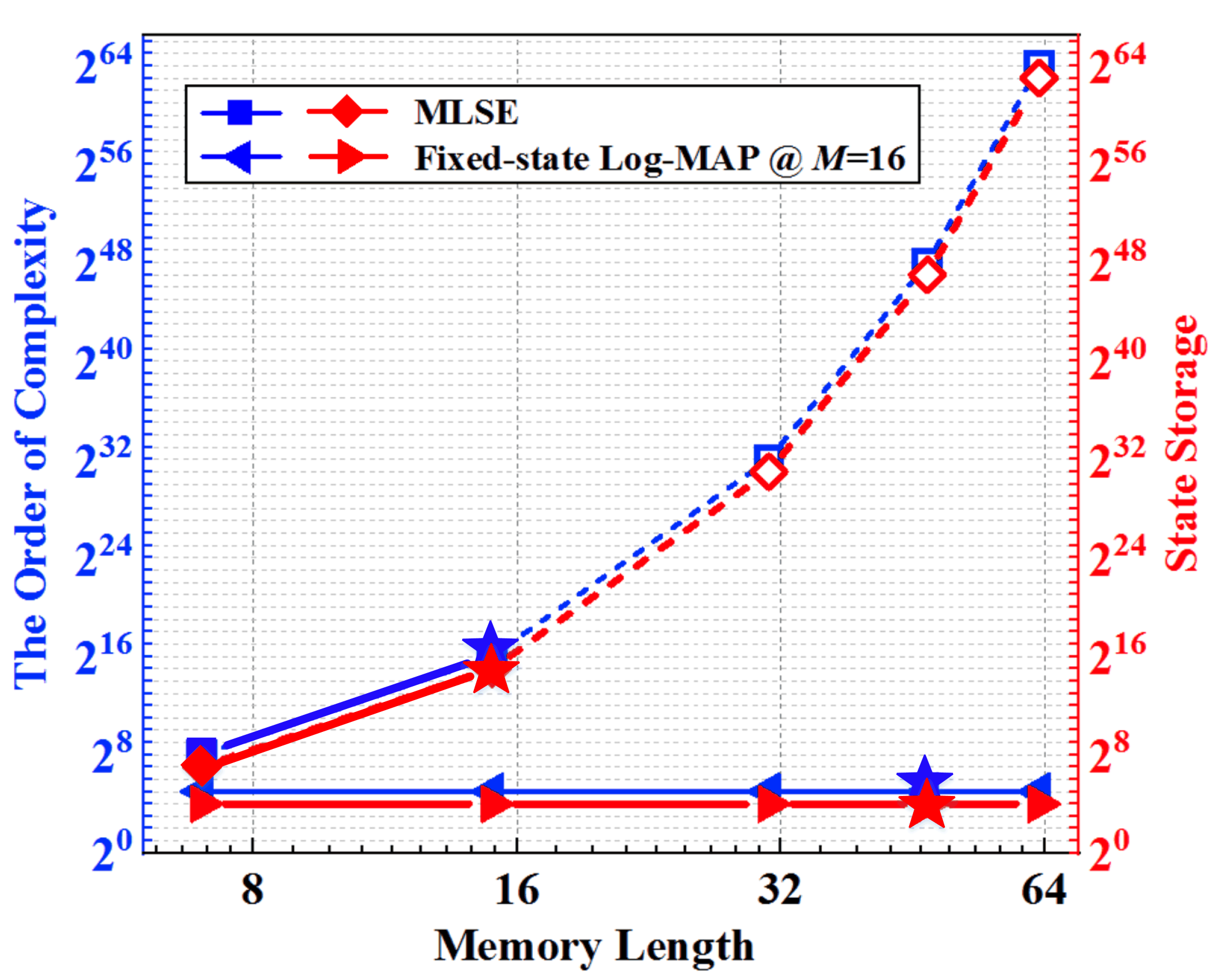}
\caption{The comparison of computational complexity and state storage between MLSE and fixed-state Log-MAP detection ($M=16$) (The solid line represents realizable ranges, and the dashed line represents the theoretical but unrealizable ranges).}
\label{Fig8}
\end{figure}

Fig. \ref{Fig8} shows a comparison of computational complexity and state storage between MLSE and fixed-state Log-MAP detection ($M=16$). The left vertical axis is the order of complexity (blue line), and the right vertical axis is the amount of state storage (red line). The square denotes MLSE, which has an exponential increase in computational complexity in order of $O(2^{L+1})$. And MLSE requires to store $2^{L}$ states. When the memory length is greater than $15$, the computing resources and internal storage space of the hardware is too overloaded to implement. The unrealizable regions are represents by dashed line. Hence, the optional maximum memory length of MLSE is 15, which is represented by pentacle. However, for the transmission links with severe distortions, the memory length is much longer than $15$. As for fixed-state Log-MAP detection denoted by triangle, the computational complexity and state storage are invariable with an increase in memory length. The computational complexity is in order of $O(2M)$, and the number of stored states is $M$. Once $M$ is fixed, the computational complexity and state storage are constant as the memory length increases. Thanks to the fixed-state Log-MAP detection independent of memory length, MLSE can be replaced by fixed-state Log-MAP detection to solve the restricted memory length. Therefore, under the same hardware condition, the fixed-state Log-MAP detection can compensate more distortions compared with MLSE.

\begin{figure}[!t]
\centering
\includegraphics[width=2.875in]{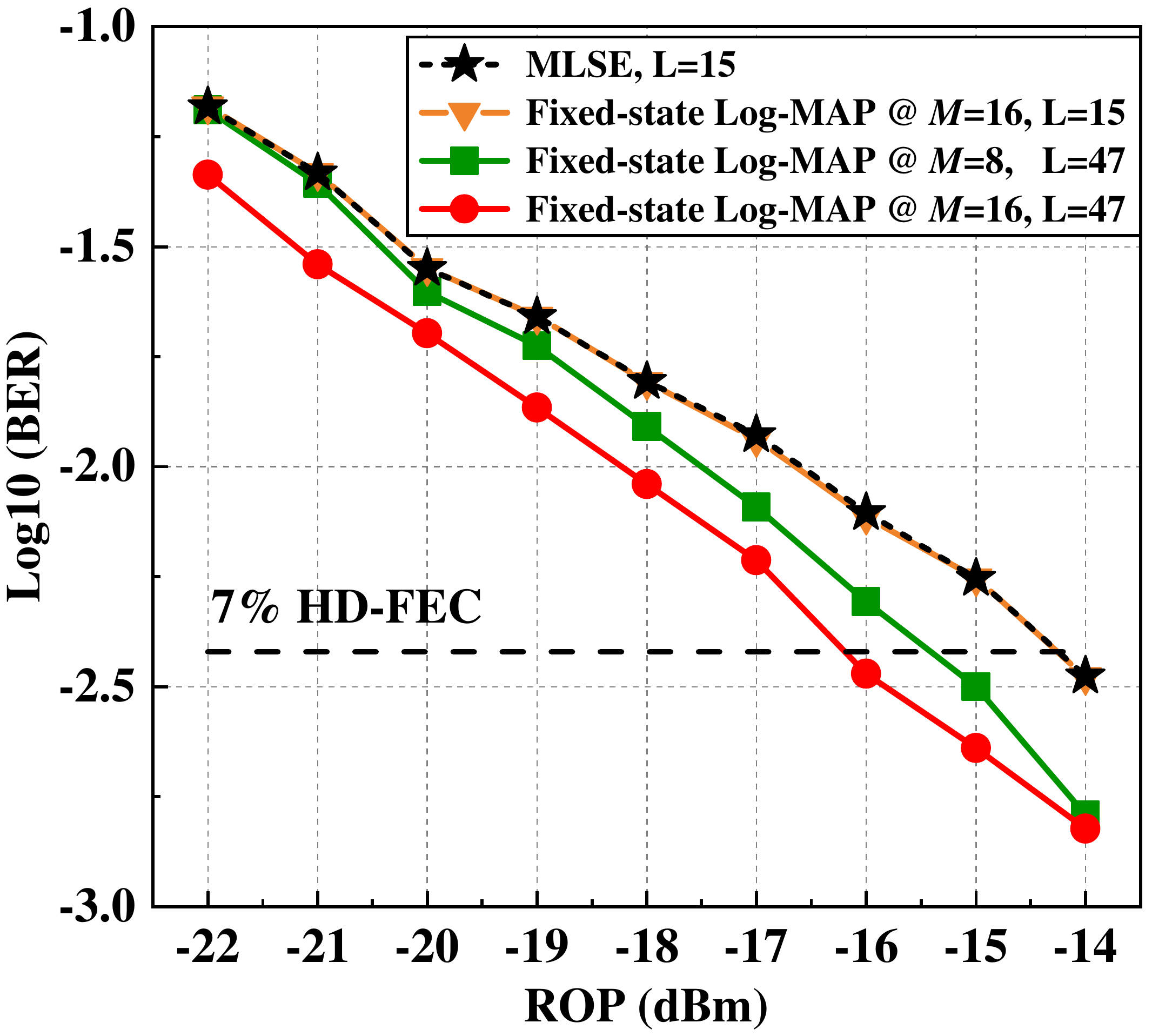}
\caption{BER performance using MLSE and fixed-state Log-MAP detection versus ROP for a 64 Gbit/s IM/DD optical OOK system over a 100 km SSMF transmission.}
\label{Fig9}
\end{figure}

Fig. \ref{Fig9} depicts the BER performance using MLSE and fixed-state Log-MAP detection versus ROP for a 64 Gbit/s IM/DD OOK system over a 100 km SSMF transmission. When the memory length is set to $L=15$, the fixed-state Log-MAP at $M=16$ has the same performance with MLSE. The ROP of -14.2 dBm is achieved at 7\% HD-FEC limit. For MLSE, the computational complexity and state storage of MLSE are $10^4$ orders of magnitude ($O(2^{L+1}=65536)$ and $2^{L}=32768$ states).  As for the fixed-state Log-MAP detection, the computational complexity and state storage of fixed-state Log-MAP detection are $10^1$ orders of magnitude ($O(2*M=32)$ and $M=16$ states). Therefore, at the same performance, the fixed-state Log-MAP detection reduces three orders of magnitude of both computational complexity and storage compared with MLSE at $L=15$. However, this is realizable best performance using MLSE with the maximum memory length of $L=15$. For a 100-km dispersion-uncompensated link, the memory length of 15 is insufficient. Theoretically, MLSE should have the same performance as fixed-state Log-MAP detection. However, owing to the huge computing and storage burden, the MLSE is unable to do anything when the memory length is greater than 15. Fortunately, the fixed-state Log-MAP overcomes the limitation of memory length. As represented in Fig. \ref{Fig4}, the memory length is set to $L=47$, which is realized by fixed-state Log-MAP detection. When the number of surviving states is $M=8$, the receiver sensitivity has a 1.3 dB improvement than that using MLSE at 7\% HD-FEC limit. However, some valid states are discarded at $M=8$. When the number of surviving states is $M=16$, the fixed-state Log-MAP detection achieves optimal performance. Compared with the optimal performance achieved by MLSE, the receiver sensitivity has a 2 dB improvement. The experimental results reveal the fixed-state Log-MAP detection can support larger memory length to compensate more distortions to realize the optimal signal detection with a low computational complexity and a less storage resources.

\section{Conclusion}
\label{section5}
In this paper, we proposed a fixed-state Log-MAP detection combined with equalizers to compensate the channel distortions for IM/DD optical systems over dispersion-uncompensated links. Compared with the classical detection such as MLSE, the computational complexity and state storage of fixed-state Log-MAP detection are independent of memory length. The fixed-state Log-MAP detection has the ability to use larger memory length to compensate more distortions. We experimentally demonstrated a C-band 64 Gbit/s IM/DD OOK system over a 100 km dispersion-uncompensated link using the fixed-state Log-MAP detection. Experimental results reveal that the fixed-state Log-MAP detection achieves better transmission performance under the same hardware condition, and the computational complexity and storage reduce by three orders of magnitude compared with MLSE. We believe this work is of significant value for future data center networks.

\ifCLASSOPTIONcaptionsoff
  \newpage
\fi

\end{document}